\documentclass{iopjournal}

\usepackage{natbib}

\begin{document}

\articletype{Letter}

\title{Identifying Arctic states vulnerable to rapid ice loss with a convolutional neural network}

\author{Matthew Patterson$^{1,2,3,*},$\orcid{0000-0002-9484-8410}, Ioana Colfescu$^{1,2,3}$\orcid{0000-0002-2819-2425}, Nathan Kirk$^{2}$\orcid{0000-0002-7989-809X} and Graeme MacGhilchrist$^{1}$\orcid{0000-0003-1409-8937}}

\affil{$^1$School of Earth and Environmental Sciences, University of St Andrews, St Andrews, UK}

\affil{$^2$School of Mathematics and Statistics, University of St Andrews, St Andrews, UK}

\affil{$^3$National Centre for Atmospheric Science, University of St Andrews, St Andrews, UK}

\affil{$^*$Corresponding author.}

\email{mrp23@st-andrews.ac.uk}

\keywords{Arctic, sea-ice, climate change, machine learning}

\begin{abstract}
This work investigates the incidence of rapid ice-loss events in the Arctic, in which a substantial fraction of September sea-ice is lost in only a few years. Such events are seen to occur in climate model simulations forced with increasing greenhouse gas concentrations, but not as yet in the relatively short observational record. We train a convolutional neural network (CNN) using six large climate model ensembles to diagnose vulnerability to rapid ice-loss events. The trained model is shown to separate higher and lower risk states in out-of-sample climate models via a leave-one-out cross-validation. The model emphasises the role of persistently thin sea-ice, remaining sea-ice extent and sustained high upper ocean heat content as predictors of rapid ice-loss. The CNN is applied to observational data over the satelite-era showing a signal for rapid ice-loss prior to a drop around 2007 and an increasing risk from around 2015 to present. Such rapid-ice loss events would likely have dramatic consequences for Arctic ecosystems, coastal communities and global shipping.
\end{abstract}

\section{Introduction}
Arctic sea-ice is projected to become consistently seasonally ice-free in this century due to anthropogenic climate change \citep{notz_observed_2016,jahn_projections_2024}, with severe impacts on Arctic ecosystems \citep{post_ecological_2013} and Arctic coastal communities \citep{huntington_societal_2022}, as well as allowing previously inaccessible Arctic waters to become navigable \citep{melia_sea_2016}. However, there remains considerable uncertainty regarding the future rate of this decline due to a combination of large internal variability, differences in climate model physics and uncertainty about future anthropogenic greenhouse gas emissions \citep{bonan_partitioning_2021}. 

 \begin{figure}
 \centering
        \includegraphics[width=\textwidth]{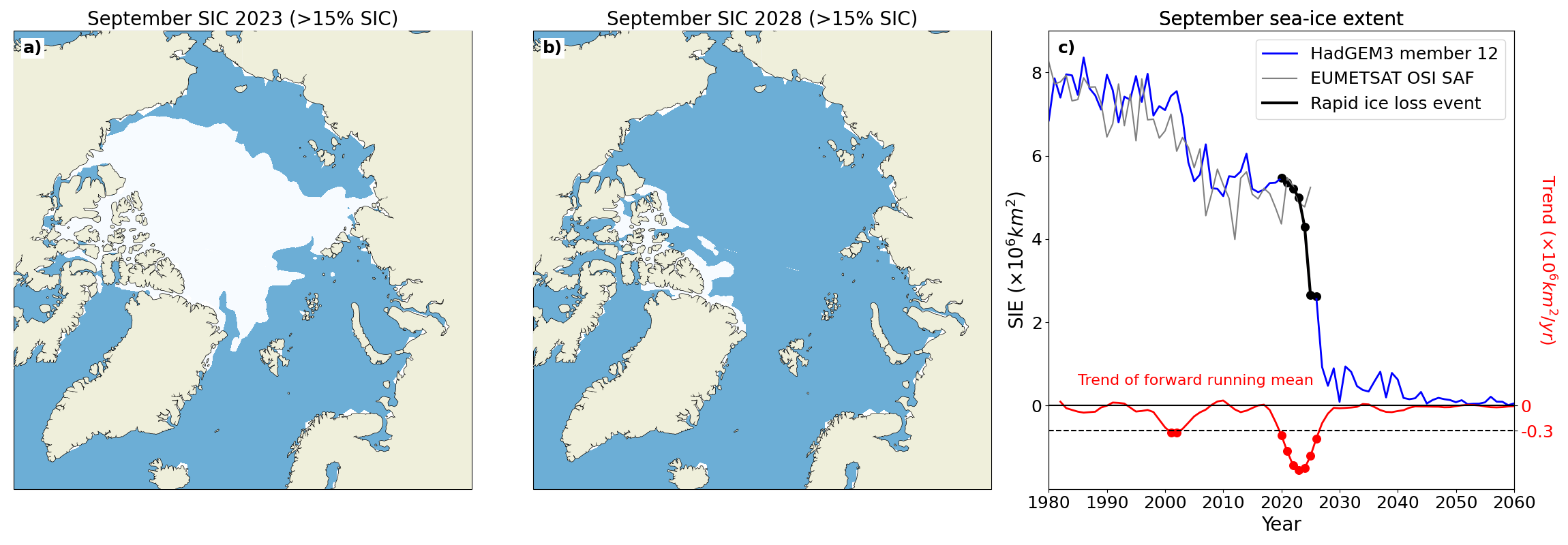}
 \caption{An example of a rapid ice-loss event in the CANARI large ensemble, member 12, run with the model HadGEM3-GC3.1-MM. Maps show regions of Arctic sea-ice concentrations above 15\% (white) in September model years a) 2023 and b) 2028. September sea-ice extent for this member (blue) is shown in c) along with detection of a rapid ice loss event (black, see main text) and the trend of the forward running-mean sea-ice extent (red). Red dots show where the forward running-mean is less than $-0.3\times10^6km^2/year$. The observational time series of sea-ice (grey) is also shown for comparison.}
\label{fig1}
\end{figure}

Climate models suggest the possibility of rapid ice-loss events (RILEs) in the Arctic, in which a large fraction of the remaining ice is lost over only a few years, exacerbating the impacts of ice-loss \citep[e.g.][]{holland_future_2006,bathiany_potential_2016,sticker_seasonality_2025}. A particularly dramatic example of a RILE, from the CANARI large ensemble \citep{schiemann_CANARI_2026}, is illustrated in figure \ref{fig1}, in which the Arctic goes from sea-ice extent of around $5\times10^6km^2$ to less than $1\times10^6km^2$ in the space of five years (figure \ref{fig1}a-c).  RILEs are distinct from the concept of tipping dynamics (abrupt and irreversible system change), which appears not to apply to summer sea-ice in either simple \citep{eisenman_nonlinear_2009,wagner_how_2015} or comprehensive models \citep{armour_reversibility_2011}. RILEs are not necessarily the result of runaway feedbacks, but rather, a consequence of both internal variability and strong externally-forced ice loss \citep{holland_future_2006}. 

Different climate models vary considerably in how frequently such events occur \citep{sticker_seasonality_2025}. Moreover, the observational record of sea-ice is relatively short and there has been no statistically significant decline in September (the annual minimum) sea-ice since 2005 \citep{england_minimal_2025}. Taking the diversity in model projections and the paucity of the observational record together, assessing the relative vulnerability of the real Arctic system to RILEs is extremely challenging. 

Machine learning is increasingly being used in climate science to find patterns in complex, multi-scale datasets, including climate models and observational datasets \citep[e.g.][]{huntingford_machine_2019,eyring_pushing_2024,bracco_machine_2025}. In the context of sea-ice, the IceNet model \citet{andersson_seasonal_2021} using a U-net architecture trained on satellite-derived sea-ice products shows improved predictive skill over state-of-the-art dynamical forecast models. On interannual timescales, \citet{hoffman_probabilistic_2025} trained neural networks to forecast multi-annual sea-ice extent anomalies from past sea-ice extent and an integrated measure of ice-thickness, finding similar performance to numerical and statistical forecasts. 

In this work, we train a convolutional neural network (CNN) to predict the state-dependent vulnerability of Arctic sea-ice extent to RILEs using data from six large climate model ensembles. The aim is to identify Arctic states which are generally vulnerable to RILEs across the multi-model ensemble and apply this trained CNN to observational data to assess the current vulnerability of the Arctic.  Data and methods are outlined in sections \ref{data} and \ref{methods}, respectively, followed by an analysis of the model performance in section \ref{validation}. This is followed by an application of the trained CNN to observational data in section \ref{observations} and then finally by discussion and conclusions in section \ref{discussion}.

\section{Data}
\label{data}
We train the CNN with six large ensembles of climate model simulations, listed in table \ref{tab1}. These models are chosen as all of the relevant variables are available and they each have at least 20 ensemble members for both historical and SSP3-7.0 scenarios. Sea-ice model comparison studies generally take one of two approaches: sub-selecting models based on their ability to capture observed trends and other metrics \citep{notz_arctic_2020} or using all available data to find more general patterns \citep{bonan_constraining_2021}. The present study is closer to the latter, as it is necessarily limited to a small number of climate models with enough members to train on. Our aim is to learn general statistical signatures of RILE vulnerability across models, hence we will show that the models chosen span a range of behaviours in terms of sea-ice decline. Moreover, the climate models use a range of sea-ice dynamical cores which will likely represent ice-ocean-atmosphere interactions in distinct ways (table \ref{tab1}). We combine the historical and SSP3-7.0 scenarios for all variables and take the years 1980-2090, focusing on the month of September only. Variables used are monthly-mean potential temperature (`thetao'),  sea-ice concentration (`siconc') and sea-ice thickness (`sithick').

\begin{table}
\caption{Climate model large ensembles used in this study along with their sea-ice models, number of ensemble members, count of RILEs and frequency of RILEs per ensemble member.}
\centering
\begin{tabular}{l c c c c}
\hline
Model & Sea-ice model & Members & RILEs & RILE freq. \\
\hline
ACCESS-ESM1-5 1 & LANL CICE4.1 &  40 & 18 & 0.45 \\
CanESM5 & LIM2 & 25 & 9 & 0.36 \\
CESM2 & CICE5.1.2 & 50 & 45 & 0.9 \\
EC-Earth3 & LIM3 & 50 & 51 & 1.02 \\
HadGEM3-GC3.1-MM & CICE-HadGEM3-GSI8  & 40 & 41 & 1.03 \\
MPI-ESM1-2-LR & MPIOM1.63  & 40 & 9 & 0.23 \\
\hline
\end{tabular}
\label{tab1}
\end{table}

For observations, we use PIOMAS SIT data \citep{schweiger_uncertainty_2011} and EUMETSAT OSI SAF sea-ice extent data \citep{lavergne_version_2019}. In-situ observations of sub-surface temperatures within the Arctic Ocean are sparse.  We use three datasets which differ considerably in their approach, to provide some understanding of the uncertainty in ocean heat content, hereafter $H$ (defined in section \ref{methods}). These include, ASTE \citep[Arctic Subpolar gyre sTate Estimate,][]{nguyen_arctic_2021}, a dynamically consistent state estimate using an adjoint method. We also use EN4 \citep{good_en4_2013}, an objective analysis of quality controlled subsurface temperature profiles, and ORAS5 \citep[Ocean Reanalysis System 5][]{zuo_ecmwf_2019}, a global eddy-permitting ocean-ice reanalysis system. For each of the $H$ datasets, we use the years 1981-2025, apart from ASTE, which spans 2002-2017.

\section{Methods}
\label{methods}

 \begin{figure}
 \centering
        \includegraphics[width=\textwidth]{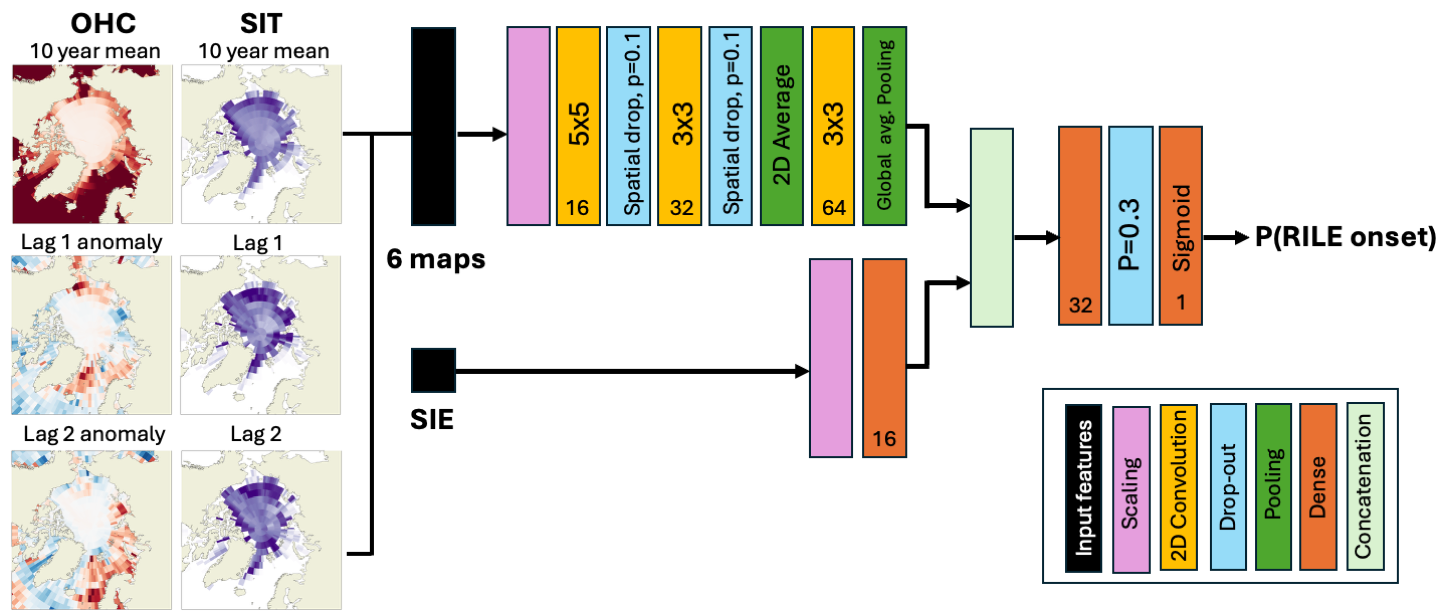}
 \caption{Architecture of the CNN used in this study. The model is fed with 6 spatial maps and 1 scalar (previous September sea-ice extent) and trained to predict the probability of the onset of a RILE. Further details of the CNN architecture are described in the main text.}
\label{fig2}
\end{figure}

\subsection{Rapid ice loss definition}
We follow the approach of \citet{auclair_role_2018} and \citet{sticker_seasonality_2025} in defining RILEs as periods when the five-year trend of the five-year running-mean of September sea-ice extent falls below $-0.3\times10^6 km^2/yr$ for at least four consecutive years. Note that we slightly modify the smoothing step to be a forward running-mean to prevent leakage of information from the future when training the model. The RILE definition is illustrated in figure \ref{fig1}c with the trend of the forward running-mean in red. A consequence of the forward running-mean is that the onset year occurs slightly before the steepest decline, while the end of the RILE is slightly before the ice finishes declining (figure \ref{fig1}c). Nevertheless, our focus is on vulnerability to the onset of RILEs rather than specifically predicting the onset timing and hence this slight shift is less of an issue. The definition is likely to be sensitive to variation in the trend threshold, running mean and requirement for at least four consecutive years to be below the threshold. However, we use this definition without variations in these parameters for consistency with previous studies \citep{auclair_role_2018, sticker_seasonality_2025}. Overall, the RILE definition serves to emphasise large, multi-year events. 

\subsection{Derived quantities, model predictors and scaling}
Ocean heat content, `$H$', is calculated as $H = c_p\int_{0}^{50}\rho (\theta-\theta_{ref})dz$, where $c_p$ is the specific heat capacity of sea water, $\rho$ is the density of sea water, $\theta$ is the ocean potential temperature, $z$ is the depth and $\theta_{ref}$ is a reference temperature, here set to $-1.8^{\circ}C$, the freezing point of Arctic water.  $H$ is calculated over the top 50m as a compromise between being deep enough to represent storage of heat and yet shallow enough to strongly affect the sea-ice. Sea-ice extent is calculated on the native grid from monthly-mean sea-ice concentration data. $H$ is also calculated on the native grid and then remapped, along with SIT, to a common $2.5^{\circ}\times2.5^{\circ}$ grid. The CNN is then provided with data for the latitudes 40N-85N.  

For most machine learning applications it is necessary to scale input variables, such that all variables are of a similar magnitude, to aid convergence in training \citep{furtado_setting_2026}. Concomitantly we wish to preserve physical thresholds when scaling and avoid our CNN simply learning climate model identity. We therefore divide sea-ice extent by the mean sea-ice extent in the climatology period, 1980-2000, for each climate model. This climatology period is chosen as it coincides with the availability of high quality observation-based data from the satelite-era. Similarly, we scale sea-ice thickness by the mean at each grid-point from the climatology period for each model. We transform $H$ as $H'=tanh(H/(3\times10^8GJm^{-2}))$ as $H$ will increase unbounded with global warming, once ice cover is lost. 


\subsection{Convolutional Neural Network and training}

We train a CNN (figure \ref{fig2}) to predict the probability that the onset of a RILE  will occur in the next five years, based on sea-ice extent and maps of SIT and $H$ in previous years. Whilst the model learns to predict the \emph{probability} of a RILE onset within the training dataset, we do not expect this to be a well-calibrated probability when applied out of sample to climate models with a different representation of relevant processes. Hence, we refer mostly to the `RILE vulnerability', with the aim that this metric is self-consistent for a given climate model in separating high and low risk states. We may therefore talk of the relative vulnerability of a climate model compared to previous states of that model but not the absolute probability of a RILE.

Specifically, we feed the CNN with maps of the backward 10-year September-mean $H'$, to represent low-frequency variability, anomalies of $H'$ with respect to this 10-year mean in the previous two Septembers, backward 10-year September-mean SIT and SIT from the previous two Septembers. There are three convolutional layers, the first of which is a 5x5 filter, followed by two 3x3 convolutions (figure \ref{fig2}). The network is designed to emphasise large-scale coherent patterns rather than small-scale signatures, which may be more model-dependent. With this goal in mind, spatial drop-out is used after the first two convolutional layers along with 2D average pooling after the second convolution and global average pooling after the third. Additionally, we feed sea-ice extent from the previous September into a dense layer which is later concatenated with the output of the convolutional layers. This concatenated layer is fed through a dense layer followed by drop-out. Apart from the final layer, which uses a sigmoid activation, all activation layers use a ReLU function. 

4 members from each large ensemble are randomly held out for testing and a further 8 for validation giving a total of 173 simulations for training. We remove any years which are part of the RILE after the onset year so that the CNN learns the precursors to RILEs and not their effects. We use a binary cross entropy loss function, defined as,

\begin{equation}
\mathcal{L}= -\frac{1}{N} \sum_{i=1}^{N} \left[ y_i \log p_i+(1-y_i)\log(1-p_i)\right]
\end{equation}

The terms, $y_i, p_i \in [0,1] $ are the $i$'th label and predicted probability, respectively, and $N$ is the number of samples. We train using an Adam optimizer with a learning rate of $2\times10^{-4}$ and a maximum of 300 epochs, using early stopping on the validation AUC after the first 100 epochs, applied with a patience of 50 epochs. To counteract the different ensemble sizes and event frequencies (table \ref{tab1}), we weight the training samples by $1/nf$, where $n$ is the number of ensemble members and $f$, the event frequency. Furthermore, given the rarity of RILEs we weight RILE events five times higher in training than non-RILEs. Results are not strongly sensitive to reasonable variations in these weightings (not shown). We repeat the training process with 10 different random seeds and show average values in all figures. Overall, the CNN architecture is relatively simple and strongly regularised, to reflect the fact that there are limited RILEs on which to train and the high temporal and spatial autocorrelation across input fields. 


 \begin{figure}
 \centering
        \includegraphics[width=\textwidth]{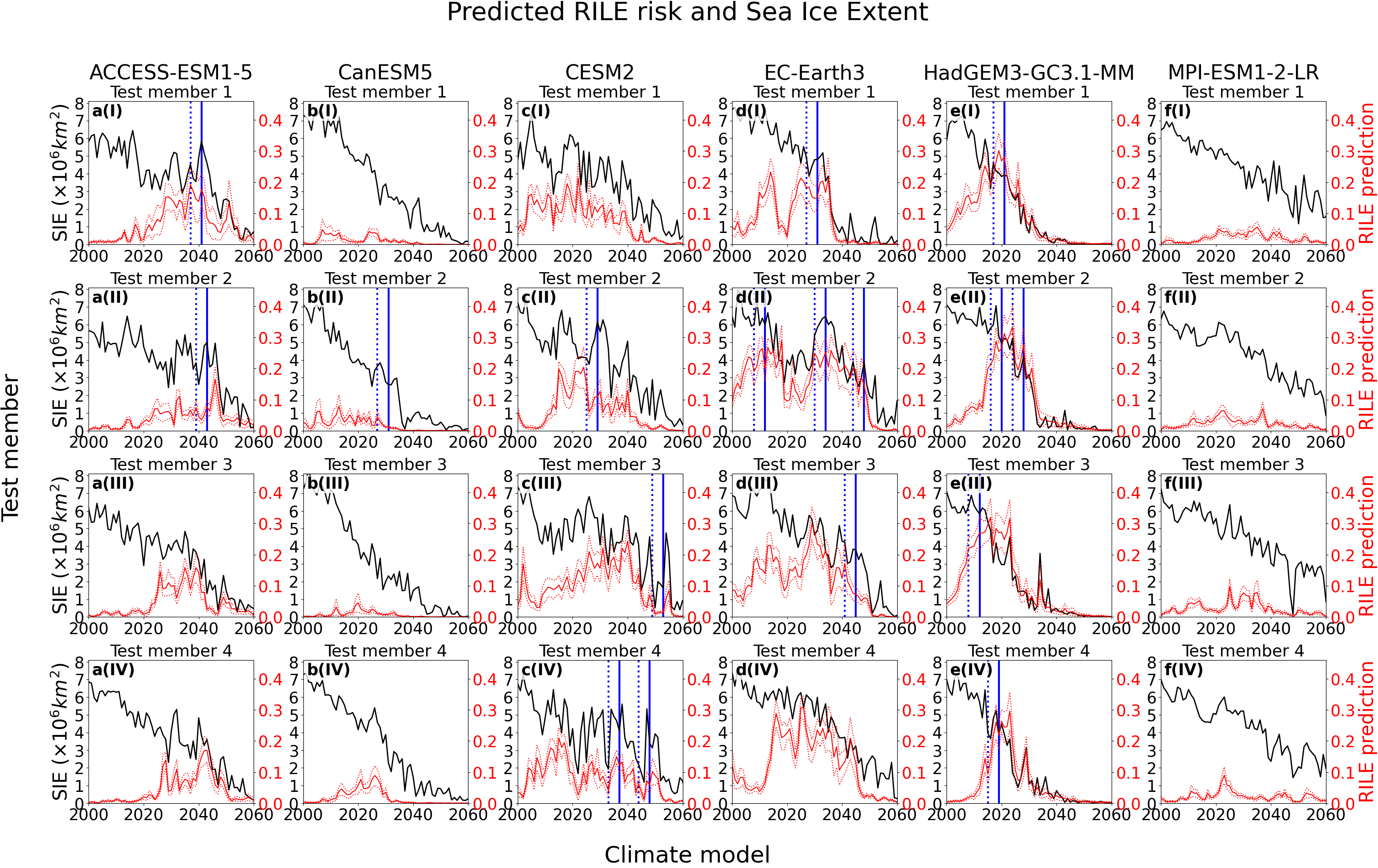}
 \caption{Predicted vulnerability to RILEs over time for simulations in the test set. Each panel shows the time series of September sea-ice extent (black) and mean predicted vulnerability (from 10 models trained with different random seeds) for RILE onset in that particular year (red, solid). The spread of these model predictions is indicated by 1 standard deviation across the models either side of the mean prediction (red, dotted). Vertical, blue solid lines indicate the actual onset years of RILEs and dotted blue lines show  when the onset is within the next five years. The columns are the six different large ensembles and rows, the four different test members. }
\label{fig3}
\end{figure}

\section{Model validation}
\label{validation}

The six climate models used to train the CNN display a diverse range of sea-ice extent variability and rates of future decline (figure \ref{fig3}, table \ref{tab1}). For example, RILEs occur around once per simulation, on average, in CESM2, EC-Earth3 and HadGEM3-GC3.1-MM, but only every four simulations for MPI-ESM1-2-LR (table \ref{tab1}). These event rates are lower than those found by \citet{sticker_seasonality_2025}, though this is likely because their study focused on the higher emissions, SSP5-8.5 scenario. Both EC-Earth3 and HadGEM3-GC3.1-MM (figure \ref{fig3}d,e) generally show rapid declines in sea-ice extent with RILEs mostly occurring in the 2020s and 2030s (supplementary figure S1). By contrast, ACCESS-ESM1-5 and CESM2 show large decadal variability, with more frequent periods of increasing sea-ice extent (figure \ref{fig3}a,c). CanESM5 and MPI-ESM1-2-LR (figure \ref{fig3}b,f) do not show very steep declines or decadal variability of the same magnitude as ACCESS-ESM1-5 or CESM2 (figure \ref{fig3}a,c) and consequently, have few RILEs.

These differing behaviours are reflected in the CNN predictions. For example, EC-Earth3 and HadGEM3-GC3.1-MM generally see higher mean RILE predictions in the test set compared to the other models (figure \ref{fig3}d,e), with RILEs generally occurring when the predicted vulnerability is elevated (e.g. figure \ref{fig3}c(I,II),d(I,II,IV) ). For these two models, predictions tend to be highest between around 2010 and 2040 (figure \ref{fig3}d,e), consistent with when the RILEs tend to occur. Predictions for ACCESS-ESM1-5 show narrow spikes, often coinciding with periods when sea-ice has recently increased (figure \ref{fig3}a(I-IV)). CESM2 shows a broader period of relatively high vulnerability (figure \ref{fig3}c(I-IV)) due to the fact that RILEs occur before 2000 and as late as the 2050s (supplementary figure S1c), likely spanning a wide range of system states. The vulnerability prediction often increases if sea-ice temporarily increases (as there is more ice to lose), for example CESM2 test member 4 (figure \ref{fig3}(IV)) or EC-Earth3 member 2 (figure \ref{fig3}d(II)).   

 CanESM5 (figure \ref{fig3}b(I-IV)) and MPI-ESM1-2-LR (figure \ref{fig3}e(I-IV)) show relatively low predictions for most years due to the lack of RILEs in these models and the CNN largely misses an event  for CanESM5 test member 2 (figure \ref{fig3}b(II)). The CNN does, however, capture elevated risk of a RILE in CanESM5 member 4, in which the sea-ice decline does not ultimately qualify as a RILE but does decline relatively rapidly (figure \ref{fig3}b(IV)). This does highlight a potential weakness of the binary nature of the RILE definition in that some large sea-ice declines are not classed as RILEs, but would likely be impactful. Note that the CNN outperforms a baseline logistic model using the same inputs flattened into a single vector (scaled Arctic-mean $H'$ and SIT, and sea-ice extent) as the simpler model relies heavily on sea-ice extent and fails to capture its non-linear relationship with vulnerability (supplementary figure S2).  
 
  \begin{figure}
 \centering
        \includegraphics[width=\textwidth]{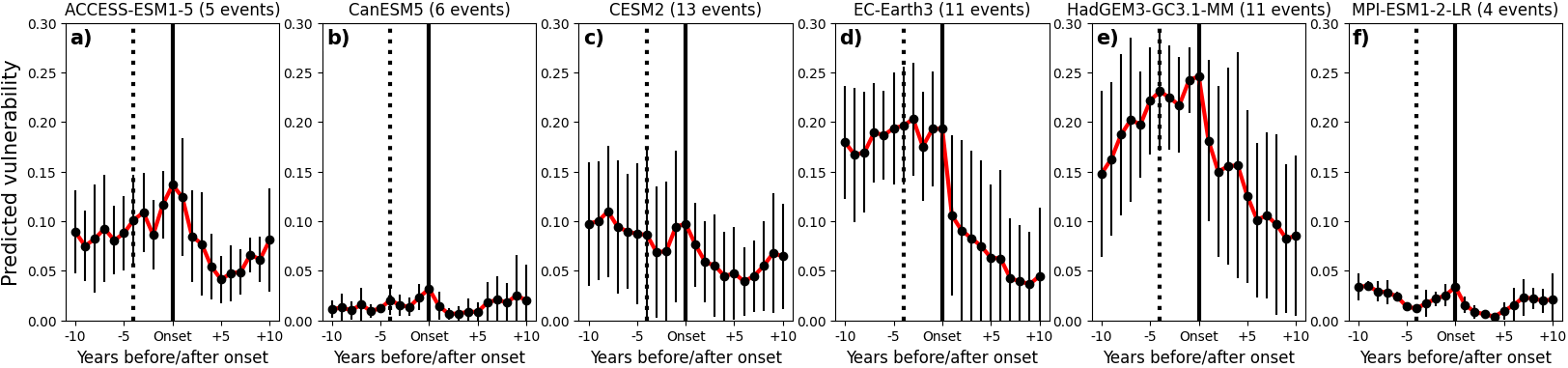}
 \caption{Composites of vulnerability predictions for each model with respect to onset timing for RILEs in the validation and test ensemble members. Shown is the mean (black dots) over all RILE events at a particular time relative to the onset timing. Error bars indicate one standard deviation either side of the mean. The solid vertical line in each panel indicates the onset year of RILEs and the dotted line indicates when onset is within the next five years.}
\label{fig3b}
\end{figure}
 
A composite analysis of RILE events in the validation and test sets, for each model, shows that predicted vulnerability is highest near the onset year for most models and declines thereafter (figure \ref{fig3b}). However, there are substantial differences between models in how far in advance such events are predicted and how well years leading to events separate from other years.  ACCESS-ESM1-5 and HadGEM3-GC3.1-MM have a clear upward trend in predicted vulnerability prior to the onset year (figure \ref{fig3b}a,e), while CanESM5 and MPI-ESM1-2-LR have a small but discernible rise over the years leading to onset (figure \ref{fig3b}b,f). CESM2 actually has a higher vulnerability 10 years before the onset, which then dips and subsequently rises to the onset year (figure \ref{fig3b}c). This may partly result from the strong decadal variability in sea-ice in CESM2 (figure \ref{fig3}c). Finally, EC-Earth3 shows flatter predicted vulnerability in the lead up to the RILE onset, possibly due to a combination of the high event frequency and broad period (and hence range of system states) over which events occur (supplementary figure S1d). In general, it is likely that RILEs are fundamentally more predictable in some climate models than others, depending on the specific underlying dynamics for RILEs in each model. Note that we combine validation and tests sets here as there are too few RILEs in just the test set, while we avoid including the training set as the CNN is directly trained on these. 

 \begin{figure}
 \centering
        \includegraphics[width=\textwidth]{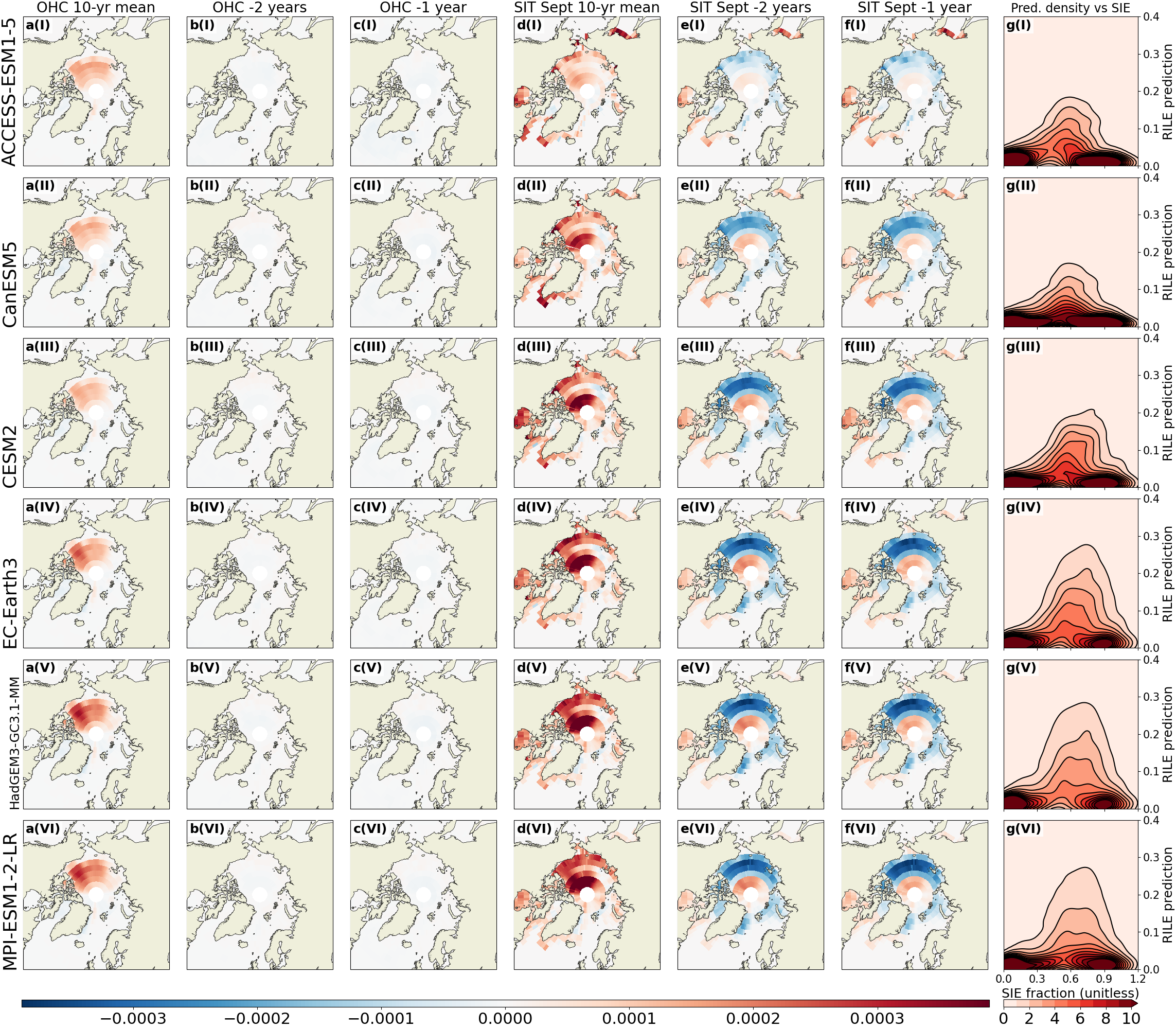}
 \caption{SHAP analysis of CNN predictions for all predictions in the test set with values above 0.001. Calculations are relative to a background of the 1980-2000 climatology period in the test set. Shown are the average contributions of the 6 input variables to predictions, relative to the background. These are a) 10-year mean $H'$, b) $H'$ anomalies at lag 2, c) $H'$ anomalies at lag 1, d) 10-year mean SIT, e) SIT at lag 2 and f) SIT at lag 1. Column g) shows a kernel density estimate of prediction values for each climate model as a function of scaled sea-ice extent.}
\label{fig4}
\end{figure}

In order to understand which variables are most important in the CNN predictions and build confidence that the CNN is learning physically sensible patterns, we use SHapley Additive Explanations \citep[SHAP,][]{lundberg_unified_2017}. SHAP is an explainable artificial intelligence technique which seeks to identify the importance of each feature in a given prediction. SHAP values will sum to the predicted value and their sign indicates whether a feature has pushed the prediction up or down, relative to a particular baseline. We separately analyse SHAP values for the six different climate models using the years 1980-2000 as the baseline. We note that SHAP relies on the assumption that features can be varied independently, which is not the case in the climate system due to the high spatial autocorrelation of climate fields like $H$ \citep{krell_influence_2025}. It is therefore likely that the signal will be blurred-out somewhat across correlated features. Nevertheless, this analysis can still give a broad idea of how the network functions.

Across the six models there is a general pattern of the 10-year running mean $H'$ raising the vulnerability to RILEs in the central Arctic and particularly in marginal seas such as the Beaufort and Chukchi Seas (figure \ref{fig4}a(I-VI)). That is, the CNN learns that high $H'$ in this region is linked to a greater chance of RILE onset. This makes sense given fact that sea-ice typically forms in the marginal seas, so additional warming here will reduce the production of sea-ice and contribute to ice-loss. Conversely, there is no relationship with seasonally uncovered regions such as the Barents and Kara seas. This is not to say that the Barents and Kara seas do not influence the ice-loss in other regions, but that $H$ and SIT values there do not contribute skill to RILE predictions. 

In spite of the importance of the 10-year mean $H'$, recent anomalies in $H'$ play effectively no role in CNN predictions (figure \ref{fig4}b(I-VI),c(I-VI) ). This suggests that it is the long-term, sustained warming of the marginal seas which contributes to predictable RILE events rather than anomalous warming in the most recent Septembers. 

The largest SHAP values occur across all models for 10-year mean SIT (figure \ref{fig4}d(I-VI)). Effectively, this means that relatively thin sea-ice, particularly in the central Arctic and on the periphery of the Canadian sector, raises the chance of RILEs. This is physically plausible given the greater susceptibility of thin ice to summer ice-loss from atmospheric variability \citep{stroeve_arctics_2012}. In contrast to the 10-year mean raising the vulnerability prediction, SIT values in the two previous years in marginal seas lower the prediction (figure \ref{fig4}e,f). This is likely because whilst thinner ice is more vulnerable, there needs to be sufficient remaining ice for a RILE to occur. If the SHAP analysis is performed for years leading to the onset of RILEs with a background of the prior to 5-10 years, similar patterns of $H$ and SIT contributing to RILE risk can be seen for EC-Earth3, HadGEM3-GC3.1-MM and MPI-ESM1-2-LR (supplementary figure S3a-f(IV,V,VI)) and to some degree, ACCESS-ESM1-5 (supplementary figure S3a-f(I)). However, CanESM5 and CESM2 differ between the two different SHAP analyses (supplementary figure S3a-f(II,III) and figure \ref{fig4}). which may be because RILEs are less clearly predicted by the CNN for these models (figure \ref{fig3b}b,c). Nevertheless, the consistency between the different SHAP maps in figure \ref{fig4} suggests that the CNN is learning meaningful precursors to RILE onset and not simply fitting to the individual climate models. 

We also plot a kernel density estimate of prediction values as a function of scaled sea-ice extent in figure \ref{fig4}g. Interestingly, high predicted vulnerability values are only seen for values of scaled sea-ice extent between 0.4 and 0.9 (that is between 40\% and 90\% of baseline sea-ice extent). This hints at a non-linear role for sea-ice extent in the predictions. This makes sense given that RILEs do not often occur during the baseline period (with the exception of CESM2 and one event in the EC-Earth3 ensemble, supplementary figure S1), nor do they occur when little ice remains. A composite difference plot for high predicted vulnerability minus low predicted vulnerability (supplementary figure S4) corroborates the findings from the SHAP analysis. That is, higher 10-year mean $H'$ (especially in the Pacific sector, supplementary figure S4a) and lower SIT (supplementary figure S4d) are associated with higher RILE risk. However, this picture only emerges when we exclude scaled sea-ice extent below 0.4 or above 0.9  (supplementary figures S4) as the low vulnerability composite mixes high and low ice states (supplementary figure S5h).


 \begin{figure}
 \centering
        \includegraphics[width=\textwidth]{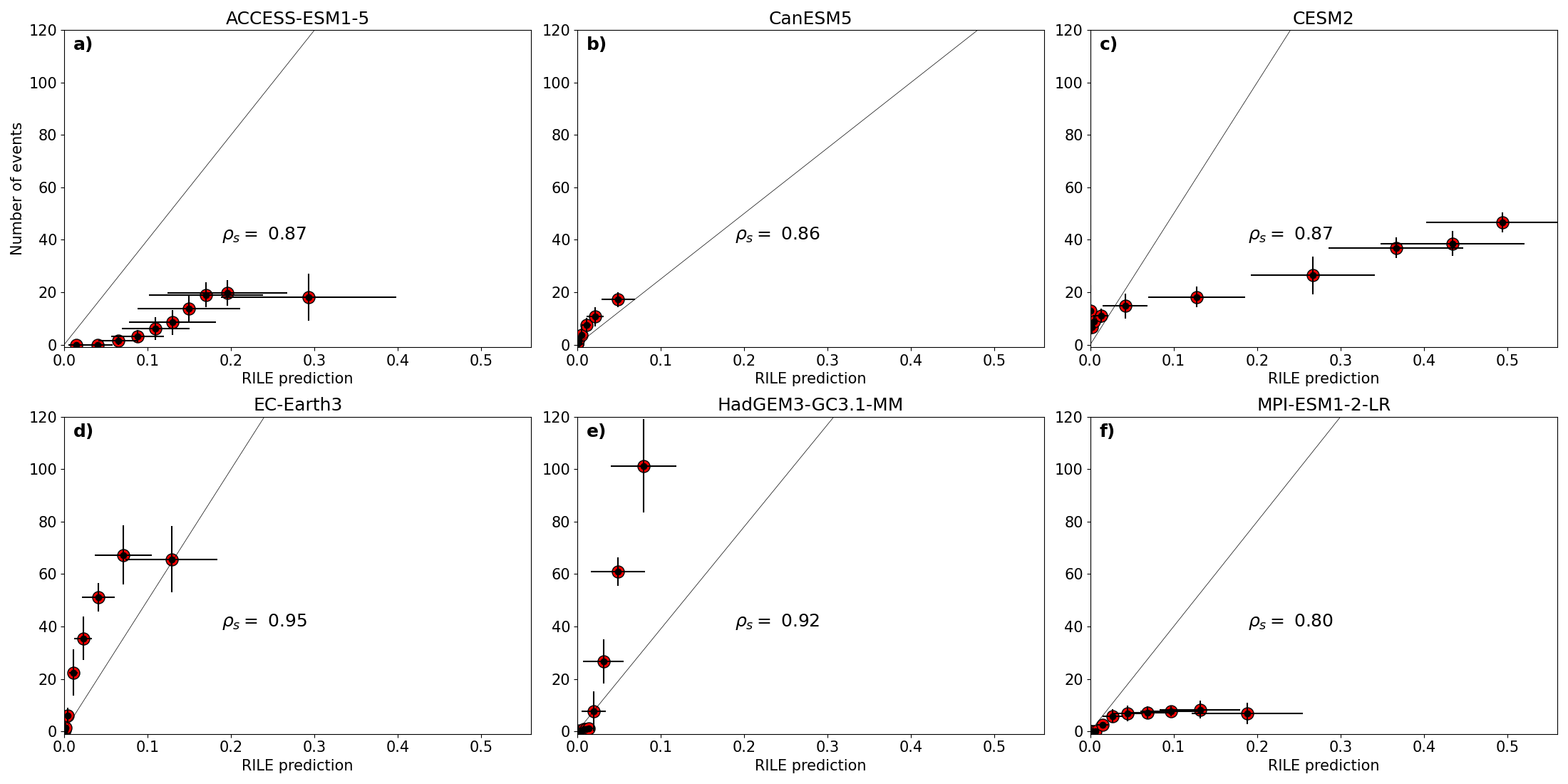}
 \caption{Out of sample predictions. In each case, the CNNs are trained with different random seeds on all but one of the models and then tested using all members of the remaining model. The figure shows the number of years with an onset in the next five years for RILE predictions binned by decile. Scatter points represent the mean of the deciles calculated from each of the 10 different CNN training runs with error bars showing the standard deviation across these deciles. The diagonal line in each panel indicates where points would lie if the CNN were a perfectly calibrated probabilistic prediction. The deciles constructed from the predictions from 10 different seeds are used to calculate the Spearman's rank coefficient (taken as a measure of monotonicity) and the mean of the Spearman's rank, calculated separately for each seed, is shown in each panel.}
\label{fig5}
\end{figure}

Our aim in this study is to apply the trained CNN to observation-based datasets to form some understanding of the current vulnerability of the Arctic to RILEs. This requires, firstly, that the CNN has learned patterns within the training data which generalize, to some degree, beyond these six climate models, and secondly, that the climate models on which the model is trained, represent the real Arctic system sufficiently well. We test the first of these requirements by performing a leave-one out cross validation \citep[e.g.][]{schlund_constraining_2020} test, removing one model, training on the remaining four and then testing with all ensemble members of the left-out model. We also exclude the left-out model from the validation set. We do not require that the left-out model is well-calibrated in terms of probability. This is because the underlying rate of RILEs varies considerably between different climate models (table \ref{tab1}) and the selected models are only a small sample of possible model representations of the Arctic system. Without a physical model or a representative historical record of RILEs, it is unlikely that it would be possible to adequately calibrate the true RILE probability. Instead, we focus on the relative vulnerability in time and require that the CNN can discriminate between high and low risk states for a given climate model. We therefore would hope to make relative statements about how much the real Arctic system's vulnerability may have increased relative to some base period.   

We bin leave-one out predictions for each climate model by the decile of the prediction (figure \ref{fig5}). This figure indicates that the number of actual RILEs generally increases with higher predictions, with Spearman's rank coefficients ranging from 0.8 to 0.95. The general monotonic increase in event occurrence with  prediction values across models suggests that the CNN is indeed able to discriminate between high and low risk states in out-of-sample situations. The diagonal lines give the expected number of events for a perfectly calibrated model, which varies slightly due to the different ensemble sizes. ACCESS-ESM1-5, CESM2 and MPI-ESM1-2-LR (figure \ref{fig5}a,c,f) scatter points lie well below this line suggesting the CNN is over-confident of events for these models, when not included in the training sample. On the other hand, when EC-Earth3 and HadGEM3-GC3.1-MM are excluded from training, the CNN tends to underestimate the vulnerability (figure \ref{fig5}d,e), while CanESM5 appears to be better calibrated (figure \ref{fig5}b). Overall it appears that the CNN can discriminate between high and low risk states, but cannot give a well calibrated estimate of RILE probabilities in absolute terms. The vulnerability predictions should therefore be used as a metric of risk relative to past states of the simulation in question. 



\section{Current Arctic vulnerability}
\label{observations}

 \begin{figure}
 \centering
        \includegraphics[width=0.6\textwidth]{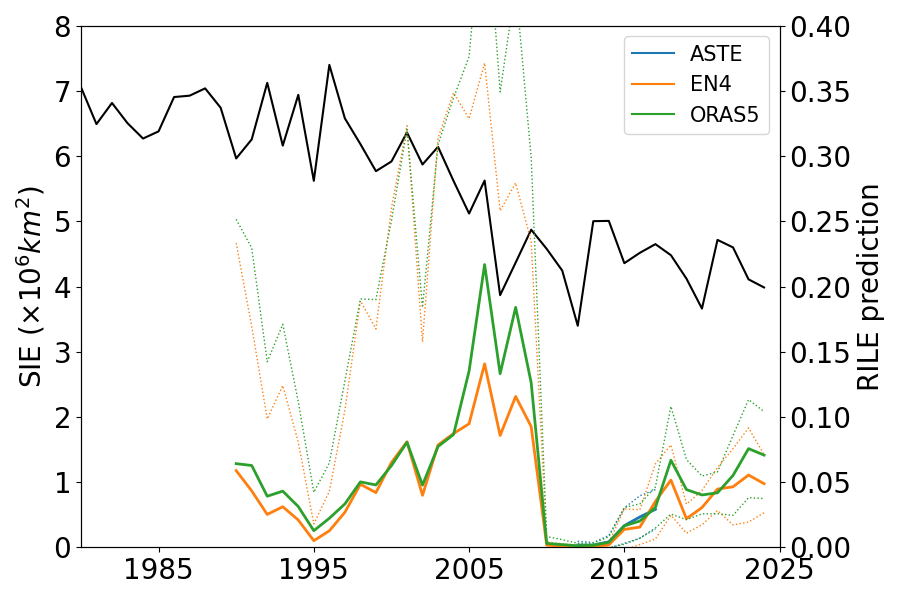}
 \caption{CNN-based predictions of relative vulnerability to Arctic RILE onset with time. Observed sea-ice time series from EUMETSAT OSI SAF is shown (black) with the mean predictions of RILE vulnerability over CNNs with training initialised from 10 random seeds shown by solid lines. Predictions are from a combination of EUMETSAT OSI SAF sea-ice extent, PIOMAS SIT and $H$ from ASTE (blue), EN4 (orange) and ORAS5 (green). Thin lines represent one standard deviation of predictions from the random seeds either side of the mean over all seeds.}
\label{fig6}
\end{figure}

We give the trained CNN data from EUMETSAT OSI SAF for sea-ice, PIOMAS for SIT and $H'$ from three different sources: ASTE, EN4  and ORAS5. Vulnerability rises from around 1995 to a peak in about 2006 (figure \ref{fig6}), a year prior to the sharp decline in summer sea-ice seen in 2007 \citep[e.g.][]{comiso_accelerated_2008,stroeve_arctic_2008}. At the time, this event contributed to concerns that summer sea-ice may decline much faster than model projections \citep{stroeve_arctic_2007, overland_when_2013}. In reality, September sea-ice extent has remained relatively stable since 2005 (figure \ref{fig6}), though such hiatus periods are often followed by accelerated loss in climate models \citep{england_minimal_2025} and there are signs that sea-ice is beginning to decline again substantially in winter \citep{chan_record-low_2026}. The 2007 decline does not qualify as a RILE under the \citet{auclair_role_2018} definition as the rapid decline did not continue over multiple years, however it is notable that the CNN detects elevated risk here. 


From 2014 onward the predictions of RILE vulnerability rise (figure \ref{fig6}). All three $H$ datasets show similar year-to-year variability. There is a slight difference in the exact values between EN4 and ORAS5 in these predictions, as ORAS5 sees a larger increase in Arctic $H$ over the period (supplementary figure S6). The ASTE dataset is relatively short (2002-2017) and the requirement for a 10-year backward mean gives only 6 years of predictions, but in that period it is consistent with the other datasets. A similar SHAP analysis to before for the prediction using ORAS5 indicates that sea-ice thinning in the Beaufort and Chukchi seas is the main driver of the increased vulnerability prediction (supplementary figure S7). It should be noted that the spread of predictions from the different random seeds is quite large (figure \ref{fig6}) compared to the predictions for in-sample climate models ((figure \ref{fig3}). This suggests that the use of different seeds is important and indicates some sensitivity in applying the CNN out-of-sample. The CNNs trained with one climate model excluded mostly show the same pattern of increased risk centred on 2005 and increasing risk from 2014 to the present day, though the details of which of these periods is deemed most vulnerable vary (supplementary figure S8).

 \begin{figure}
 \centering
        \includegraphics[width=0.8\textwidth]{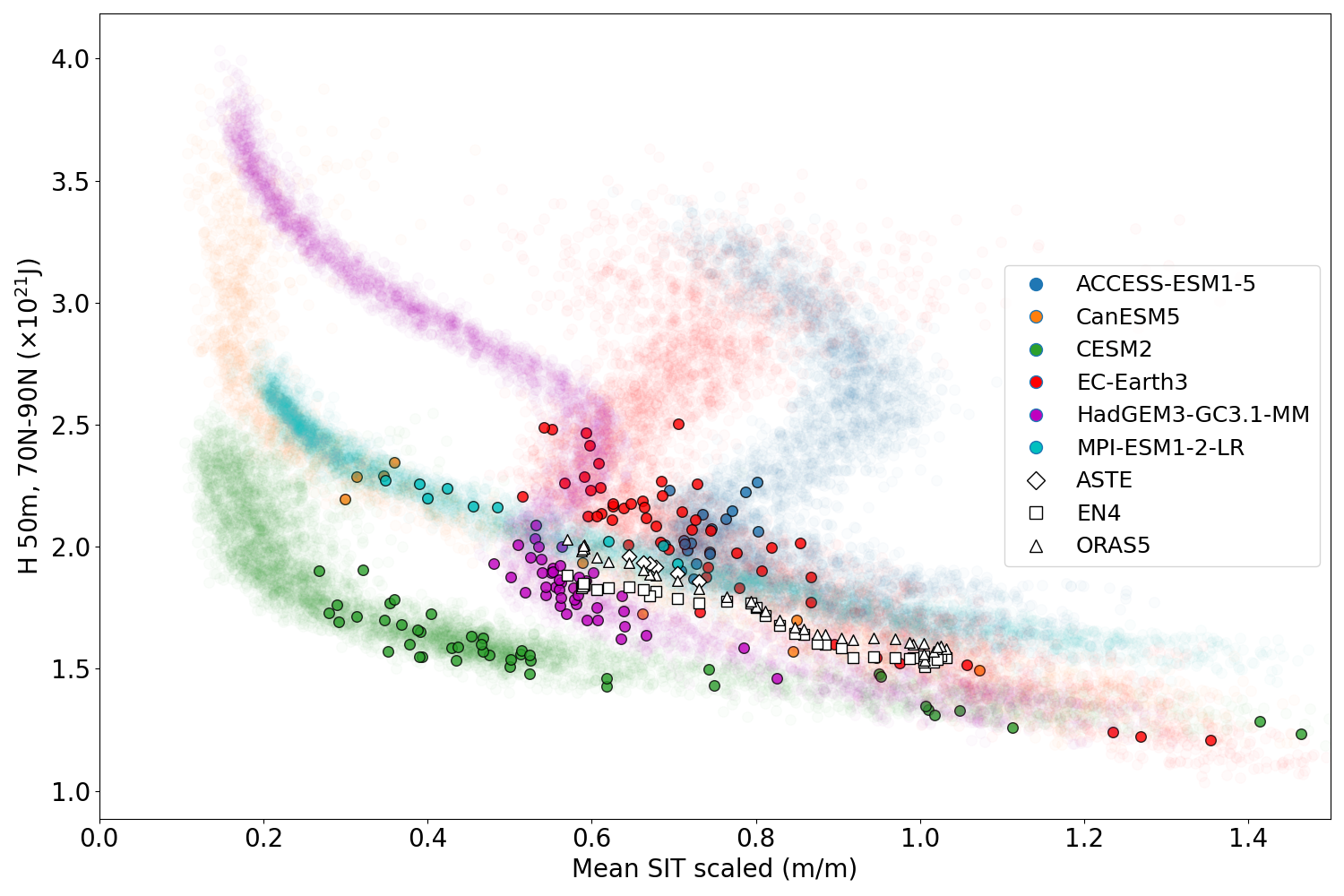}
 \caption{Low frequency $H$ vs scaled SIT state space. Backward 10-year mean $H$ is plotted against backward 10 year-mean SIT (scaled by the 1980-2000 climatology)  for climate models for all times 1980-2090 in all climate models in the study with RILE onset indicated by darker coloured points. Observational values are shown for PIOMAS SIT with the ASTE (2002-2017, diamonds), EN4 (1980-2024, squares) and ORAS5 (1980-2024, triangles) datasets. }
\label{fig7}
\end{figure}

Motivated by the strong role of backward 10-year mean $H$ and SIT in explaining CNN predictions of RILE vulnerability the climate models (figure \ref{fig4}), we plot backward 10-year Arctic $H$ vs scaled SIT in figure \ref{fig7} along with RILE onsets. We scale SIT to better reflect what the CNN actually sees by dividing the Arctic-mean SIT by the mean in the climatology period (1980-2000) for each climate model. Note that the $H$ and scaled SIT values are backward means and so are the precursors to RILEs and not contemporaneous. Each model occupies a slightly different region of the phase space, but with a general progression from high SIT and low $H$ to lower SIT and rapidly rising $H$. For some climate models there is a turning point, where mean SIT increases again with increasing $H$. This may be because of a large amount of thinner marginal sea-ice being lost, leaving the remaining Arctic sea-ice thicker on average. For ACCESS-ESM1-5 (blue), HadGEM3-GC3.1-MM (pink) and to a lesser degree, EC-Earth3 (red), this turning point is where RILEs typically occur. 

Interestingly, there is a large cluster of RILEs (or imminent RILEs) preceded by scaled SIT values of around 0.4 to 0.8 compared to the climatology period (figure \ref{fig7}). The cluster is bounded by Arctic $H$ values around $2.6\times10^{21}J$. A similar clustering is also seen if additional climate models are added to the figure (supplementary figure S9). This clustering is consistent with the fact that RILEs only occur for scaled sea-ice extent values between 40\% and 90\% of those in the climatology period (figure \ref{fig4}g). That is, there is a `transition zone' between high and low ice states where almost all RILEs occur.  

Plotting observation-based values from PIOMAS and the $H$ datasets, it can be seen that the Arctic system has moved to scaled SIT values around 0.6, in a similar region of state space to the cluster of RILEs in climate models (figure \ref{fig7}). Whilst this relationship is purely statistical, this coincidence between $H$ and SIT conditions for model RILEs and the state of the observed system, underscores the findings in figure \ref{fig6} that the Arctic system is becoming increasingly vulnerable to RILEs.


\section{Discussion and conclusions}
\label{discussion}

This study trains a Convolutional Neural Network (CNN, figure \ref{fig2}) using climate model data to learn the state-dependent vulnerability of the Arctic to rapid ice loss events (RILEs). The model is able to discriminate between high risk and low risk Arctic states based on maps of ocean heat content in the top 50m ($H$) and sea-ice thickness (SIT) in recent years (figure \ref{fig5}). The CNN learns a pattern whereby higher long-term $H$ and ice-thinning act to raise the predicted vulnerability to RILEs, whereas recent SIT maps inform the network of whether or not sufficient ice remains for RILEs to occur (figure \ref{fig4}). Applying the CNN to observational data suggests that the real world is moving towards a region of state space where RILEs more frequently occur in climate models (figures \ref{fig6}, \ref{fig7}).

Being statistical in nature and trained using only a subset of climate models without direct reference to observational constraints, our approach is unable to attach an absolute, calibrated probability to predictions of vulnerability to RILEs. Rather, our network is trained to discriminate in a relative sense between high and low risk states. Moreover, our predictions for the real Arctic system are limited by observational uncertainty, given the paucity of in situ observations and short observational period. Nevertheless, the use of deep learning techniques allows us to identify consistent precursor patterns to rare events across a large dataset. 

Our study has assumed that RILEs are rare, stochastic events which are not necessarily predictable, but which can become more likely when a system enters certain states. In line with previous research, we highlight the role of low sea-ice thickness as a pre-conditioner for such events \citep[e.g.][]{sticker_seasonality_2025}. It is also possible that other physical processes may contribute to the predictability of RILEs such as enhanced ocean heat transport \citep{oldenburg_respective_2024}, sustained atmospheric forcing driving ice export \citep{sumata_regime_2023}, increased atmospheric heat transport \citep{wang_recent_2025} or break-up of ice following increased storminess \citep{kohout_storm-induced_2014}. The most important drivers of RILEs may vary considerably between different climate models and further research is required to establish which processes may contribute most in the real world to inform more detailed assessment and early warning systems for Arctic sea-ice vulnerability.


%


\funding{MP and IC were funded through the National Centre for Atmospheric Science and University of St Andrews via the NCAS St Andrews partnership. GM received funding through a UKRI future leaders fellowship (MR/W013835/1).}

\roles{MP conceived of the study, performed the analysis and wrote the original draft. IC provided supervision and IC, NK, GM and MP contributed to review and editing of the manuscript.}

\data{CMIP6 data can be downloaded from the Earth System Grid Federation \url{https://esgf-ui.ceda.ac.uk/search} and CESM data is available via  \url{https://gdex.ucar.edu/datasets/d651056/dataaccess/}. PIOMAS data is available at \url{https://pscfiles.apl.uw.edu/zhang/PIOMAS/data/v2.1/heff/} and EUMETSAT OSI SAF from \url{https://osisaf-hl.met.no/archive/osisaf/sea-ice-index/v3p0/timeseries/}. For ocean heat content, ORAS5 is available at the climate data store \url{https://cds.climate.copernicus.eu/datasets/reanalysis-oras5?tab=overview}, ASTE from \url{https://arcticdata.io/catalog/portals/ASTE} and EN4 from \url{https://www.metoffice.gov.uk/hadobs/en4/download.html}.}


\bibliographystyle{plainnat}
\bibliography{library}

@article{bracco_machine_2025,
	title = {Machine learning for the physics of climate},
	volume = {7},
	copyright = {2024 Springer Nature Limited},
	issn = {2522-5820},
	url = {https://www.nature.com/articles/s42254-024-00776-3},
	doi = {10.1038/s42254-024-00776-3},
	language = {en},
	number = {1},
	urldate = {2025-04-08},
	journal = {Nature Reviews Physics},
	publisher = {Nature Publishing Group},
	author = {Bracco, Annalisa and Brajard, Julien and Dijkstra, Henk A. and Hassanzadeh, Pedram and Lessig, Christian and Monteleoni, Claire},
	month = jan,
	year = {2025},
	pages = {6--20},
}

@article{jahn_projections_2024,
	title = {Projections of an ice-free {Arctic} {Ocean}},
	volume = {5},
	copyright = {2024 Springer Nature Limited},
	issn = {2662-138X},
	url = {https://www.nature.com/articles/s43017-023-00515-9},
	doi = {10.1038/s43017-023-00515-9},
	language = {en},
	number = {3},
	urldate = {2025-04-24},
	journal = {Nature Reviews Earth \& Environment},
	publisher = {Nature Publishing Group},
	author = {Jahn, Alexandra and Holland, Marika M. and Kay, Jennifer E.},
	month = mar,
	year = {2024},
	pages = {164--176},
}

@article{notz_arctic_2020,
	title = {Arctic {Sea} {Ice} in {CMIP6}},
	volume = {47},
	copyright = {©2020. The Authors.},
	issn = {1944-8007},
	url = {https://onlinelibrary.wiley.com/doi/abs/10.1029/2019GL086749},
	doi = {10.1029/2019GL086749},
	language = {en},
	number = {10},
	urldate = {2025-07-15},
	journal = {Geophysical Research Letters},
	author = {Notz, Dirk and Community, Simip},
	year = {2020},
	note = {\_eprint: https://agupubs.onlinelibrary.wiley.com/doi/pdf/10.1029/2019GL086749},
	pages = {e2019GL086749},
}

@article{andersson_seasonal_2021,
	title = {Seasonal {Arctic} sea ice forecasting with probabilistic deep learning},
	volume = {12},
	copyright = {2021 The Author(s)},
	issn = {2041-1723},
	url = {https://www.nature.com/articles/s41467-021-25257-4},
	doi = {10.1038/s41467-021-25257-4},
	language = {en},
	number = {1},
	urldate = {2025-09-03},
	journal = {Nature Communications},
	publisher = {Nature Publishing Group},
	author = {Andersson, Tom R. and Hosking, J. Scott and Pérez-Ortiz, María and Paige, Brooks and Elliott, Andrew and Russell, Chris and Law, Stephen and Jones, Daniel C. and Wilkinson, Jeremy and Phillips, Tony and Byrne, James and Tietsche, Steffen and Sarojini, Beena Balan and Blanchard-Wrigglesworth, Eduardo and Aksenov, Yevgeny and Downie, Rod and Shuckburgh, Emily},
	month = aug,
	year = {2021},
	pages = {5124},
}

@article{wang_recent_2025,
	title = {Recent slowing of {Arctic} sea ice melt tied to multidecadal {NAO} variability},
	volume = {16},
	copyright = {2025 The Author(s)},
	issn = {2041-1723},
	url = {https://www.nature.com/articles/s41467-025-63520-0},
	doi = {10.1038/s41467-025-63520-0},
	language = {en},
	number = {1},
	urldate = {2025-09-29},
	journal = {Nature Communications},
	publisher = {Nature Publishing Group},
	author = {Wang, Cen and Su, Hui and Zhai, Chengxing and Zheng, Jianqiu and Yu, Shiwei and Mo, Huisi and Wang, Yanjia and Jiang, Linwei},
	month = sep,
	year = {2025},
	pages = {8504},
}

@article{eisenman_nonlinear_2009,
	title = {Nonlinear threshold behavior during the loss of {Arctic} sea ice},
	volume = {106},
	url = {https://www.pnas.org/doi/abs/10.1073/pnas.0806887106},
	doi = {10.1073/pnas.0806887106},
	number = {1},
	urldate = {2026-07-17},
	journal = {Proceedings of the National Academy of Sciences},
	publisher = {Proceedings of the National Academy of Sciences},
	author = {Eisenman, I. and Wettlaufer, J. S.},
	month = jan,
	year = {2009},
	pages = {28--32},
}

@article{sticker_seasonality_2025,
	title = {Seasonality and scenario dependence of rapid {Arctic} sea ice loss events in {CMIP6} simulations},
	volume = {19},
	issn = {1994-0416},
	url = {https://tc.copernicus.org/articles/19/3259/2025/},
	doi = {10.5194/tc-19-3259-2025},
	language = {English},
	number = {8},
	urldate = {2026-07-17},
	journal = {The Cryosphere},
	publisher = {Copernicus GmbH},
	author = {Sticker, Annelies and Massonnet, François and Fichefet, Thierry and DeRepentigny, Patricia and Jahn, Alexandra and Docquier, David and Wyburn-Powell, Christopher and Quint, Daphne and Shivers, Erica and Ortiz, Makayla},
	month = aug,
	year = {2025},
	pages = {3259--3277},
}

@article{notz_observed_2016,
	title = {Observed {Arctic} sea-ice loss directly follows anthropogenic {CO2} emission},
	volume = {354},
	url = {https://www.science.org/doi/full/10.1126/science.aag2345},
	doi = {10.1126/science.aag2345},
	number = {6313},
	urldate = {2026-07-17},
	journal = {Science},
	publisher = {American Association for the Advancement of Science},
	author = {Notz, Dirk and Stroeve, Julienne},
	month = nov,
	year = {2016},
	pages = {747--750},
}

@article{post_ecological_2013,
	title = {Ecological {Consequences} of {Sea}-{Ice} {Decline}},
	volume = {341},
	url = {https://www.science.org/doi/full/10.1126/science.1235225},
	doi = {10.1126/science.1235225},
	number = {6145},
	urldate = {2026-07-17},
	journal = {Science},
	publisher = {American Association for the Advancement of Science},
	author = {Post, Eric and Bhatt, Uma S. and Bitz, Cecilia M. and Brodie, Jedediah F. and Fulton, Tara L. and Hebblewhite, Mark and Kerby, Jeffrey and Kutz, Susan J. and Stirling, Ian and Walker, Donald A.},
	month = aug,
	year = {2013},
	pages = {519--524},
}

@article{melia_sea_2016,
	title = {Sea ice decline and 21st century trans-{Arctic} shipping routes},
	volume = {43},
	copyright = {©2016. The Authors.},
	issn = {1944-8007},
	url = {https://onlinelibrary.wiley.com/doi/abs/10.1002/2016GL069315},
	doi = {10.1002/2016GL069315},
	language = {en},
	number = {18},
	urldate = {2026-07-17},
	journal = {Geophysical Research Letters},
	author = {Melia, N. and Haines, K. and Hawkins, E.},
	year = {2016},
	note = {\_eprint: https://agupubs.onlinelibrary.wiley.com/doi/pdf/10.1002/2016GL069315},
	pages = {9720--9728},
}

@article{huntington_societal_2022,
	title = {Societal implications of a changing {Arctic} {Ocean}},
	volume = {51},
	issn = {1654-7209},
	url = {https://doi.org/10.1007/s13280-021-01601-2},
	doi = {10.1007/s13280-021-01601-2},
	language = {en},
	number = {2},
	urldate = {2026-07-17},
	journal = {Ambio},
	author = {Huntington, Henry P. and Zagorsky, Andrey and Kaltenborn, Bjørn P. and Shin, Hyoung Chul and Dawson, Jackie and Lukin, Maija and Dahl, Parnuna Egede and Guo, Peiqing and Thomas, David N.},
	month = feb,
	year = {2022},
	pages = {298--306},
}

@article{bonan_partitioning_2021,
	title = {Partitioning uncertainty in projections of {Arctic} sea ice},
	volume = {16},
	issn = {1748-9326},
	url = {https://doi.org/10.1088/1748-9326/abe0ec},
	doi = {10.1088/1748-9326/abe0ec},
	language = {en},
	number = {4},
	urldate = {2026-07-17},
	journal = {Environmental Research Letters},
	publisher = {IOP Publishing},
	author = {Bonan, David B and Lehner, Flavio and Holland, Marika M},
	month = mar,
	year = {2021},
	pages = {044002},
}

@article{holland_future_2006,
	title = {Future abrupt reductions in the summer {Arctic} sea ice},
	volume = {33},
	copyright = {Copyright 2006 by the American Geophysical Union.},
	issn = {1944-8007},
	url = {https://onlinelibrary.wiley.com/doi/abs/10.1029/2006GL028024},
	doi = {10.1029/2006GL028024},
	language = {en},
	number = {23},
	urldate = {2026-07-17},
	journal = {Geophysical Research Letters},
	author = {Holland, Marika M. and Bitz, Cecilia M. and Tremblay, Bruno},
	year = {2006},
	note = {\_eprint: https://agupubs.onlinelibrary.wiley.com/doi/pdf/10.1029/2006GL028024},
}

@article{bathiany_potential_2016,
	chapter = {Journal of Climate},
	title = {On the {Potential} for {Abrupt} {Arctic} {Winter} {Sea} {Ice} {Loss}},
	volume = {29},
	issn = {0894-8755, 1520-0442},
	url = {https://journals.ametsoc.org/view/journals/clim/29/7/jcli-d-15-0466.1.xml},
	doi = {10.1175/JCLI-D-15-0466.1},
	language = {EN},
	number = {7},
	urldate = {2026-07-17},
	journal = {Journal of Climate},
	publisher = {American Meteorological Society},
	author = {Bathiany, S. and Notz, D. and Mauritsen, T. and Raedel, G. and Brovkin, V.},
	month = apr,
	year = {2016},
	pages = {2703--2719},
}

@article{hoffman_probabilistic_2025,
	title = {Probabilistic {Forecasts} of {September} {Arctic} {Sea} {Ice} {Extent} at the {Interannual} {Timescale} {With} {Data}-{Driven} {Statistical} {Models}},
	volume = {2},
	copyright = {© 2025 The Author(s). Journal of Geophysical Research: Machine Learning and Computation published by Wiley Periodicals LLC on behalf of American Geophysical Union.},
	issn = {2993-5210},
	url = {https://onlinelibrary.wiley.com/doi/abs/10.1029/2025JH000669},
	doi = {10.1029/2025JH000669},
	language = {en},
	number = {3},
	urldate = {2026-07-17},
	journal = {Journal of Geophysical Research: Machine Learning and Computation},
	author = {Hoffman, L. and Massonnet, F. and Sticker, A.},
	year = {2025},
	note = {\_eprint: https://agupubs.onlinelibrary.wiley.com/doi/pdf/10.1029/2025JH000669},
	pages = {e2025JH000669},
}

@article{england_minimal_2025,
	title = {Minimal {Arctic} {Sea} {Ice} {Loss} in the {Last} 20 {Years}, {Consistent} {With} {Internal} {Climate} {Variability}},
	volume = {52},
	copyright = {© 2025. The Author(s).},
	issn = {1944-8007},
	url = {https://onlinelibrary.wiley.com/doi/abs/10.1029/2025GL116175},
	doi = {10.1029/2025GL116175},
	language = {en},
	number = {15},
	urldate = {2026-07-17},
	journal = {Geophysical Research Letters},
	author = {England, M. R. and Polvani, L. M. and Screen, J. and Chan, A. C.},
	year = {2025},
	note = {\_eprint: https://agupubs.onlinelibrary.wiley.com/doi/pdf/10.1029/2025GL116175},
	pages = {e2025GL116175},
}

@article{auclair_role_2018,
	title = {The {Role} of {Ocean} {Heat} {Transport} in {Rapid} {Sea} {Ice} {Declines} in the {Community} {Earth} {System} {Model} {Large} {Ensemble}},
	volume = {123},
	copyright = {©2018. American Geophysical Union. All Rights Reserved.},
	issn = {2169-9291},
	url = {https://onlinelibrary.wiley.com/doi/abs/10.1029/2018JC014525},
	doi = {10.1029/2018JC014525},
	language = {en},
	number = {12},
	urldate = {2026-07-17},
	journal = {Journal of Geophysical Research: Oceans},
	author = {Auclair, Gabriel and Tremblay, L. Bruno},
	year = {2018},
	note = {\_eprint: https://agupubs.onlinelibrary.wiley.com/doi/pdf/10.1029/2018JC014525},
	pages = {8941--8957},
}

@article{eyring_pushing_2024,
	title = {Pushing the frontiers in climate modelling and analysis with machine learning},
	volume = {14},
	copyright = {2024 Springer Nature Limited},
	issn = {1758-6798},
	url = {https://www.nature.com/articles/s41558-024-02095-y},
	doi = {10.1038/s41558-024-02095-y},
	language = {en},
	number = {9},
	urldate = {2026-07-17},
	journal = {Nature Climate Change},
	publisher = {Nature Publishing Group},
	author = {Eyring, Veronika and Collins, William D. and Gentine, Pierre and Barnes, Elizabeth A. and Barreiro, Marcelo and Beucler, Tom and Bocquet, Marc and Bretherton, Christopher S. and Christensen, Hannah M. and Dagon, Katherine and Gagne, David John and Hall, David and Hammerling, Dorit and Hoyer, Stephan and Iglesias-Suarez, Fernando and Lopez-Gomez, Ignacio and McGraw, Marie C. and Meehl, Gerald A. and Molina, Maria J. and Monteleoni, Claire and Mueller, Juliane and Pritchard, Michael S. and Rolnick, David and Runge, Jakob and Stier, Philip and Watt-Meyer, Oliver and Weigel, Katja and Yu, Rose and Zanna, Laure},
	month = sep,
	year = {2024},
	pages = {916--928},
}

@article{huntingford_machine_2019,
	title = {Machine learning and artificial intelligence to aid climate change research and preparedness},
	volume = {14},
	issn = {1748-9326},
	url = {https://doi.org/10.1088/1748-9326/ab4e55},
	doi = {10.1088/1748-9326/ab4e55},
	language = {en},
	number = {12},
	urldate = {2026-07-17},
	journal = {Environmental Research Letters},
	publisher = {IOP Publishing},
	author = {Huntingford, Chris and Jeffers, Elizabeth S and Bonsall, Michael B and Christensen, Hannah M and Lees, Thomas and Yang, Hui},
	month = nov,
	year = {2019},
	pages = {124007},
}

@article{wagner_how_2015,
	chapter = {Journal of Climate},
	title = {How {Climate} {Model} {Complexity} {Influences} {Sea} {Ice} {Stability}},
	volume = {28},
	issn = {0894-8755, 1520-0442},
	url = {https://journals.ametsoc.org/view/journals/clim/28/10/jcli-d-14-00654.1.xml},
	doi = {10.1175/JCLI-D-14-00654.1},
	language = {EN},
	number = {10},
	urldate = {2026-07-17},
	journal = {Journal of Climate},
	publisher = {American Meteorological Society},
	author = {Wagner, Till J. W. and Eisenman, Ian},
	month = may,
	year = {2015},
	pages = {3998--4014},
}

@article{armour_reversibility_2011,
	title = {The reversibility of sea ice loss in a state-of-the-art climate model},
	volume = {38},
	copyright = {Copyright 2011 by the American Geophysical Union.},
	issn = {1944-8007},
	url = {https://onlinelibrary.wiley.com/doi/abs/10.1029/2011GL048739},
	doi = {10.1029/2011GL048739},
	language = {en},
	number = {16},
	urldate = {2026-07-17},
	journal = {Geophysical Research Letters},
	author = {Armour, K. C. and Eisenman, I. and Blanchard-Wrigglesworth, E. and McCusker, K. E. and Bitz, C. M.},
	year = {2011},
	note = {\_eprint: https://agupubs.onlinelibrary.wiley.com/doi/pdf/10.1029/2011GL048739},
}

@article{bonan_constraining_2021,
	title = {Constraining the {Date} of a {Seasonally} {Ice}-{Free} {Arctic} {Using} a {Simple} {Model}},
	volume = {48},
	copyright = {© 2021. American Geophysical Union. All Rights Reserved.},
	issn = {1944-8007},
	url = {https://onlinelibrary.wiley.com/doi/abs/10.1029/2021GL094309},
	doi = {10.1029/2021GL094309},
	language = {en},
	number = {18},
	urldate = {2026-07-20},
	journal = {Geophysical Research Letters},
	author = {Bonan, David B. and Schneider, Tapio and Eisenman, Ian and Wills, Robert C. J.},
	year = {2021},
	note = {\_eprint: https://agupubs.onlinelibrary.wiley.com/doi/pdf/10.1029/2021GL094309},
	pages = {e2021GL094309},
}

@article{schweiger_uncertainty_2011,
	title = {Uncertainty in modeled {Arctic} sea ice volume},
	volume = {116},
	copyright = {Copyright 2011 by the American Geophysical Union.},
	issn = {2156-2202},
	url = {https://onlinelibrary.wiley.com/doi/abs/10.1029/2011JC007084},
	doi = {10.1029/2011JC007084},
	language = {en},
	number = {C8},
	urldate = {2026-07-20},
	journal = {Journal of Geophysical Research: Oceans},
	author = {Schweiger, Axel and Lindsay, Ron and Zhang, Jinlun and Steele, Mike and Stern, Harry and Kwok, Ron},
	year = {2011},
	note = {\_eprint: https://agupubs.onlinelibrary.wiley.com/doi/pdf/10.1029/2011JC007084},
}

@article{good_en4_2013,
	title = {{EN4}: {Quality} controlled ocean temperature and salinity profiles and monthly objective analyses with uncertainty estimates},
	volume = {118},
	copyright = {©2013. American Geophysical Union. All Rights Reserved.},
	issn = {2169-9291},
	shorttitle = {{EN4}},
	url = {https://onlinelibrary.wiley.com/doi/abs/10.1002/2013JC009067},
	doi = {10.1002/2013JC009067},
	language = {en},
	number = {12},
	urldate = {2026-07-20},
	journal = {Journal of Geophysical Research: Oceans},
	author = {Good, Simon A. and Martin, Matthew J. and Rayner, Nick A.},
	year = {2013},
	note = {\_eprint: https://agupubs.onlinelibrary.wiley.com/doi/pdf/10.1002/2013JC009067},
	pages = {6704--6716},
}

@article{nguyen_arctic_2021,
	title = {The {Arctic} {Subpolar} {Gyre} {sTate} {Estimate}: {Description} and {Assessment} of a {Data}-{Constrained}, {Dynamically} {Consistent} {Ocean}-{Sea} {Ice} {Estimate} for 2002–2017},
	volume = {13},
	copyright = {© 2021. The Authors. Journal of Advances in Modeling Earth Systems published by Wiley Periodicals LLC on behalf of American Geophysical Union.},
	issn = {1942-2466},
	shorttitle = {The {Arctic} {Subpolar} {Gyre} {sTate} {Estimate}},
	url = {https://onlinelibrary.wiley.com/doi/abs/10.1029/2020MS002398},
	doi = {10.1029/2020MS002398},
	language = {en},
	number = {5},
	urldate = {2026-07-20},
	journal = {Journal of Advances in Modeling Earth Systems},
	author = {Nguyen, An T. and Pillar, Helen and Ocaña, Victor and Bigdeli, Arash and Smith, Timothy A. and Heimbach, Patrick},
	year = {2021},
	note = {\_eprint: https://agupubs.onlinelibrary.wiley.com/doi/pdf/10.1029/2020MS002398},
	pages = {e2020MS002398},
}

@article{zuo_ecmwf_2019,
	title = {The {ECMWF} operational ensemble reanalysis–analysis system for ocean and sea ice: a description of the system and assessment},
	volume = {15},
	issn = {1812-0784},
	shorttitle = {The {ECMWF} operational ensemble reanalysis–analysis system for ocean and sea ice},
	url = {https://os.copernicus.org/articles/15/779/2019/},
	doi = {10.5194/os-15-779-2019},
	language = {English},
	number = {3},
	urldate = {2026-07-20},
	journal = {Ocean Science},
	publisher = {Copernicus GmbH},
	author = {Zuo, Hao and Balmaseda, Magdalena Alonso and Tietsche, Steffen and Mogensen, Kristian and Mayer, Michael},
	month = jun,
	year = {2019},
	pages = {779--808},
}

@article{lavergne_version_2019,
	title = {Version 2 of the {EUMETSAT} {OSI} {SAF} and {ESA} {CCI} sea-ice concentration climate data records},
	volume = {13},
	issn = {1994-0416},
	url = {https://tc.copernicus.org/articles/13/49/2019/},
	doi = {10.5194/tc-13-49-2019},
	language = {English},
	number = {1},
	urldate = {2026-07-20},
	journal = {The Cryosphere},
	publisher = {Copernicus GmbH},
	author = {Lavergne, Thomas and Sørensen, Atle Macdonald and Kern, Stefan and Tonboe, Rasmus and Notz, Dirk and Aaboe, Signe and Bell, Louisa and Dybkjær, Gorm and Eastwood, Steinar and Gabarro, Carolina and Heygster, Georg and Killie, Mari Anne and Brandt Kreiner, Matilde and Lavelle, John and Saldo, Roberto and Sandven, Stein and Pedersen, Leif Toudal},
	month = jan,
	year = {2019},
	pages = {49--78},
}

@article{furtado_setting_2026,
	chapter = {Bulletin of the American Meteorological Society},
	title = {Setting the {Standard}: {Recommended} {Practices} for {Data} {Preprocessing} in {Data}-{Driven} {Climate} {Prediction}},
	volume = {107},
	issn = {0003-0007, 1520-0477},
	shorttitle = {Setting the {Standard}},
	url = {https://journals.ametsoc.org/view/journals/bams/107/6/BAMS-D-24-0292.1.xml},
	doi = {10.1175/BAMS-D-24-0292.1},
	language = {EN},
	number = {6},
	urldate = {2026-07-21},
	journal = {Bulletin of the American Meteorological Society},
	publisher = {American Meteorological Society},
	author = {Furtado, Jason C. and Molina, Maria J. and Arcodia, Marybeth C. and Anderson, Weston and Beucler, Tom and Callahan, John A. and Ciasto, Laura M. and Gensini, Vittorio A. and L’Heureux, Michelle and Pegion, Kathleen and Pérez-Carrasquilla, Jhayron S. and Sonnewald, Maike and Takahashi, Ken and Xiang, Baoqiang and Zimmerman, Brian G.},
	month = jun,
	year = {2026},
	pages = {E1386--E1401},
}

@article{chan_record-low_2026,
	title = {Record-low 2025 and 2026 ice extents restore {Arctic} winter sea-ice decline},
	volume = {123},
	url = {https://www.pnas.org/doi/10.1073/pnas.2614134123},
	doi = {10.1073/pnas.2614134123},
	number = {30},
	urldate = {2026-07-22},
	journal = {Proceedings of the National Academy of Sciences},
	publisher = {Proceedings of the National Academy of Sciences},
	author = {Chan, Duo and Silvano, Alessandro and Josey, Simon A.},
	month = jul,
	year = {2026},
	pages = {e2614134123},
}

@inproceedings{lundberg_unified_2017,
	address = {Red Hook, NY, USA},
	series = {{NIPS}'17},
	title = {A unified approach to interpreting model predictions},
	isbn = {978-1-5108-6096-4},
	url = {https://dl.acm.org/doi/10.5555/3295222.3295230},
	urldate = {2026-07-27},
	booktitle = {Proceedings of the 31st {International} {Conference} on {Neural} {Information} {Processing} {Systems}},
	publisher = {Curran Associates Inc.},
	author = {Lundberg, Scott M. and Lee, Su-In},
	month = dec,
	year = {2017},
	pages = {4768--4777},
}

@article{stroeve_arctics_2012,
	title = {The {Arctic}’s rapidly shrinking sea ice cover: a research synthesis},
	volume = {110},
	issn = {1573-1480},
	shorttitle = {The {Arctic}’s rapidly shrinking sea ice cover},
	url = {https://doi.org/10.1007/s10584-011-0101-1},
	doi = {10.1007/s10584-011-0101-1},
	language = {en},
	number = {3},
	urldate = {2026-07-27},
	journal = {Climatic Change},
	author = {Stroeve, Julienne C. and Serreze, Mark C. and Holland, Marika M. and Kay, Jennifer E. and Malanik, James and Barrett, Andrew P.},
	month = feb,
	year = {2012},
	pages = {1005--1027},
}

@article{schlund_constraining_2020,
	title = {Constraining {Uncertainty} in {Projected} {Gross} {Primary} {Production} {With} {Machine} {Learning}},
	volume = {125},
	copyright = {©2020. The Authors.},
	issn = {2169-8961},
	url = {https://onlinelibrary.wiley.com/doi/abs/10.1029/2019JG005619},
	doi = {10.1029/2019JG005619},
	language = {en},
	number = {11},
	urldate = {2026-07-27},
	journal = {Journal of Geophysical Research: Biogeosciences},
	author = {Schlund, Manuel and Eyring, Veronika and Camps-Valls, Gustau and Friedlingstein, Pierre and Gentine, Pierre and Reichstein, Markus},
	year = {2020},
	note = {\_eprint: https://agupubs.onlinelibrary.wiley.com/doi/pdf/10.1029/2019JG005619},
	pages = {e2019JG005619},
}

@article{comiso_accelerated_2008,
	title = {Accelerated decline in the {Arctic} sea ice cover},
	volume = {35},
	copyright = {Copyright 2008 by the American Geophysical Union.},
	issn = {1944-8007},
	url = {https://onlinelibrary.wiley.com/doi/abs/10.1029/2007GL031972},
	doi = {10.1029/2007GL031972},
	language = {en},
	number = {1},
	urldate = {2026-07-28},
	journal = {Geophysical Research Letters},
	author = {Comiso, Josefino C. and Parkinson, Claire L. and Gersten, Robert and Stock, Larry},
	year = {2008},
	note = {\_eprint: https://agupubs.onlinelibrary.wiley.com/doi/pdf/10.1029/2007GL031972},
}

@article{stroeve_arctic_2008,
	title = {Arctic {Sea} {Ice} {Extent} {Plummets} in 2007},
	volume = {89},
	copyright = {©2008. American Geophysical Union. All Rights Reserved.},
	issn = {2324-9250},
	url = {https://onlinelibrary.wiley.com/doi/abs/10.1029/2008EO020001},
	doi = {10.1029/2008EO020001},
	language = {en},
	number = {2},
	urldate = {2026-07-28},
	journal = {Eos, Transactions American Geophysical Union},
	author = {Stroeve, Julienne and Serreze, Mark and Drobot, Sheldon and Gearheard, Shari and Holland, Marika and Maslanik, James and Meier, Walt and Scambos, Ted},
	year = {2008},
	note = {\_eprint: https://agupubs.onlinelibrary.wiley.com/doi/pdf/10.1029/2008EO020001},
	pages = {13--14},
}

@article{stroeve_arctic_2007,
	title = {Arctic sea ice decline: {Faster} than forecast},
	volume = {34},
	copyright = {Copyright 2007 by the American Geophysical Union.},
	issn = {1944-8007},
	shorttitle = {Arctic sea ice decline},
	url = {https://onlinelibrary.wiley.com/doi/abs/10.1029/2007GL029703},
	doi = {10.1029/2007GL029703},
	language = {en},
	number = {9},
	urldate = {2026-07-28},
	journal = {Geophysical Research Letters},
	author = {Stroeve, Julienne and Holland, Marika M. and Meier, Walt and Scambos, Ted and Serreze, Mark},
	year = {2007},
	note = {\_eprint: https://agupubs.onlinelibrary.wiley.com/doi/pdf/10.1029/2007GL029703},
}

@article{overland_when_2013,
	title = {When will the summer {Arctic} be nearly sea ice free?},
	volume = {40},
	copyright = {©2013. American Geophysical Union. All Rights Reserved.},
	issn = {1944-8007},
	url = {https://onlinelibrary.wiley.com/doi/abs/10.1002/grl.50316},
	doi = {10.1002/grl.50316},
	language = {en},
	number = {10},
	urldate = {2026-07-28},
	journal = {Geophysical Research Letters},
	author = {Overland, James E. and Wang, Muyin},
	year = {2013},
	note = {\_eprint: https://agupubs.onlinelibrary.wiley.com/doi/pdf/10.1002/grl.50316},
	pages = {2097--2101},
}

@article{sumata_regime_2023,
	title = {Regime shift in {Arctic} {Ocean} sea ice thickness},
	volume = {615},
	copyright = {2023 The Author(s)},
	issn = {1476-4687},
	url = {https://www.nature.com/articles/s41586-022-05686-x},
	doi = {10.1038/s41586-022-05686-x},
	language = {en},
	number = {7952},
	urldate = {2026-07-28},
	journal = {Nature},
	publisher = {Nature Publishing Group},
	author = {Sumata, Hiroshi and de Steur, Laura and Divine, Dmitry V. and Granskog, Mats A. and Gerland, Sebastian},
	month = mar,
	year = {2023},
	pages = {443--449},
}

@article{kohout_storm-induced_2014,
	title = {Storm-induced sea-ice breakup and the implications for ice extent},
	volume = {509},
	copyright = {2014 Springer Nature Limited},
	issn = {1476-4687},
	url = {https://www.nature.com/articles/nature13262},
	doi = {10.1038/nature13262},
	language = {en},
	number = {7502},
	urldate = {2026-07-28},
	journal = {Nature},
	publisher = {Nature Publishing Group},
	author = {Kohout, A. L. and Williams, M. J. M. and Dean, S. M. and Meylan, M. H.},
	month = may,
	year = {2014},
	pages = {604--607},
}

@article{oldenburg_respective_2024,
	chapter = {Journal of Climate},
	title = {The {Respective} {Roles} of {Ocean} {Heat} {Transport} and {Surface} {Heat} {Fluxes} in {Driving} {Arctic} {Ocean} {Warming} and {Sea} {Ice} {Decline}},
	volume = {37},
	issn = {0894-8755, 1520-0442},
	url = {https://journals.ametsoc.org/view/journals/clim/37/4/JCLI-D-23-0399.1.xml},
	doi = {10.1175/JCLI-D-23-0399.1},
	language = {EN},
	number = {4},
	urldate = {2026-07-28},
	journal = {Journal of Climate},
	publisher = {American Meteorological Society},
	author = {Oldenburg, Dylan and Kwon, Young-Oh and Frankignoul, Claude and Danabasoglu, Gokhan and Yeager, Stephen and Kim, Who M.},
	month = jan,
	year = {2024},
	pages = {1431--1448},
}

@article{schiemann_canari_2026,
	title = {The {CANARI} {HadGEM3} {Large} {Ensemble}: {Design}, and evaluation of historical experiments},
	shorttitle = {The {CANARI} {HadGEM3} {Large} {Ensemble}},
	url = {https://egusphere.copernicus.org/preprints/2026/egusphere-2026-2281/},
	doi = {10.5194/egusphere-2026-2281},
	language = {English},
	urldate = {2026-09-10},
	journal = {EGUsphere},
	publisher = {Copernicus GmbH},
	author = {Schiemann, Reinhard K. H. and Lister, Grenville and Hatcher, Rosalyn and Hodson, Dan and Lawrence, Bryan and Shaffrey, Len and Cole, Jeff and Dittus, Andrea and Mecking, Jenny and Robson, Jon and Wilson, Simon and Aksenov, Yevgeny and Blaker, Adam T. and Harvey, Ben and Martínez-Alvarado, Oscar and Osprey, Annette and Rynders, Stephanie and Ahmadi, Sharar and Aylmer, Jake and Baker, Laura and Case, David and Gentile, Emanuele Silvio and George, Steve and Hodges, Kevin and Karlowska, Eliza and Lang, Charlotte and Lu, Hua and O'Callaghan, Niamh and Osmolska, Weronika and Osprey, Scott and Phillips, Tony and Schröder, David and Smith, Robin and Turner, Andrew G. and Woolnough, Steve and Andrews, Martin and Coward, Andrew and Hardiman, Steven and Harle, James and Rostron, John W. and Sexton, David M. H. and Sinha, Bablu and Williams, Jonny and Wilson, Chris},
	month = aug,
	year = {2026},
	pages = {1--33},
}

@article{krell_influence_2025,
	title = {The influence of correlated features on neural network attribution methods in geoscience},
	volume = {4},
	issn = {2634-4602},
	url = {https://www.cambridge.org/core/journals/environmental-data-science/article/influence-of-correlated-features-on-neural-network-attribution-methods-in-geoscience/3CFA2387032AF4BDEFB2C00C55E31886#article},
	doi = {10.1017/eds.2025.19},
	language = {en},
	urldate = {2026-09-11},
	journal = {Environmental Data Science},
	author = {Krell, Evan and Mamalakis, Antonios and King, Scott A. and Tissot, Philippe and Ebert-Uphoff, Imme},
	month = jan,
	year = {2025},
	pages = {e29},
}

\end{document}